\documentclass[prl,twocolumn,nofootinbib]{revtex4-1}
\usepackage{graphicx}
\usepackage{color}
\usepackage{dcolumn}
\usepackage{bm}
\usepackage{amssymb}
\usepackage[bottom]{footmisc}
\usepackage{footnote}
\usepackage{lipsum}
\usepackage{wrapfig}
\usepackage{sidecap}
\renewcommand{\thefootnote}{\fnsymbol{footnote}}

\newcommand\blfootnote[1]{%
\begingroup
\renewcommand\thefootnote{}\footnote{#1}%
\addtocounter{footnote}{-1}%
\endgroup}
\begin{document}
\title{Gate tunable giant anisotropic resistance in ultra-thin GaTe}
\author{Hanwen Wang,$^{1,2*}$ Mao-Lin Chen,$^{1,2*}$ Mengjian Zhu,$^{3*}$ Yaning Wang,$^{1,2*}$ Baojuan Dong,$^{1,2}$ Xingdan Sun,$^{1,2}$ Xiaorong Zhang,$^{4,6}$ Shimin Cao,$^{7,9}$ Xiaoxi Li,$^{1,2}$ Jianqi Huang,$^{1,2}$ Lei Zhang,$^{1,2}$ Weilai Liu,$^{1,2}$ Dongming Sun,$^{1,2}$ Yu Ye,$^{8,9}$ Teng Yang,$^{1,2\dagger}$ Huaihong Guo,$^{10}$ Chengbing Qin,$^{4,6\dagger}$ Liantuan Xiao,$^{4,6}$ Jing Zhang,$^{5,6}$ Jianhao Chen,$^{7,9\dagger}$ Zheng Vitto Han,$^{1,2,5\dagger}$ and Zhidong Zhang$^{1,2}$}

\affiliation{$^{1}$Shenyang National Laboratory for Materials Science, Institute of Metal Research, Chinese Academy of Sciences, Shenyang 110016, China}
\affiliation{$^{2}$School of Material Science and Engineering, University of Science and Technology of China, Anhui 230026, China}

\affiliation{$^{3}$College of Advanced Interdisciplinary Studies, National University of Defense Technology, Changsha 410073, People's Republic of China}
\affiliation{$^{4}$State Key Laboratory of Quantum Optics and Quantum Optics Devices, Institute of Laser Spectroscopy, Shanxi University, Taiyuan 030006, P. R. China}
\affiliation{$^{5}$State Key Laboratory of Quantum Optics and Quantum Optics Devices, Institute of Opto-Electronics, Shanxi University, Taiyuan 030006, P. R. China}
\affiliation{$^{6}$Collaborative Innovation Center of Extreme Optics, Shanxi University, Taiyuan 030006, P.R.China}
\affiliation{$^{7}$International Center for Quantum Materials, School of Physics, Peking University, Beijing 100871, PR China}
\affiliation{$^{8}$State Key Lab for Mesoscopic Physics and School of Physics, Peking University, Beijing, China}
\affiliation{$^{9}$Collaborative Innovation Center of Quantum Matter, Beijing 100871, PR China}
\affiliation{$^{10}$College of Sciences, Liaoning Shihua University, Fushun, 113001, China}


\maketitle
\blfootnote{\textup{*} These authors contribute equally.}

\blfootnote{$^\dagger$Corresponding to: Yangteng@imr.ac.cn, Chbqin@sxu.edu.cn, jhChen@pku.edu.cn, and vitto.han@gmail.com}

\textbf{In crystals, the duplication of atoms often follows different periodicity along different directions. It thus gives rise to the so called anisotropy, which is usually even more pronounced in two dimensional (2D) materials due to the absence of $\textbf{\textit{z}}$ dimension. Indeed, in the emerging 2D materials, electrical anisotropy has been one of the focuses in recent experimental efforts.\cite{Fengnian_Xia_NC_2014, BAs_AM2018, Xiaolong_Xu_ACSAMI_2017, Yinan_Liu_Adv_Mater_2018, Liang_Li_ACSNano_2017, MiaoFeng_NC_2015, Liang_Li_AdvMater_2018, Shengxue_Yang_AFM_2018} However, key understandings of the in-plane anisotropic resistance in low-symmetry 2D materials, as well as demonstrations of model devices taking advantage of it, have proven difficult. Here, we show that, in few-layered semiconducting GaTe, electrical conductivity along $\textbf{\textit{x}}$ and $\textbf{\textit{y}}$ directions of the 2D crystal can be gate tuned from a ratio of less than one order to as large as 10$^{3}$. This effect is further demonstrated to yield an anisotropic memory resistor behaviour in ultra-thin GaTe, when equipped with an architecture of van der Waals floating gate. Our findings of gate tunable giant anisotropic resistance (GAR) effect pave the way for potential applications in nano-electronics such as multifunctional directional memories in the 2D limit.}

 \bigskip
 \bigskip

It is known that when electrically measuring a bulk material, the resulted conductivity may manifest strong directional dependencies.\cite{Nature1939, MgB2} Discrepancy of conductivity along different in-plane directions in layered bulk crystals can sometimes be as large as a few hundreds, which however often requires a certain conditions such as the presence of large external magnetic field.\cite{WTe2_JWang} Recently, low-symmetry two dimensional (2D) materials have attracted significant attentions because of the potential applications of in-plane anisotropic nanoelectronics.\cite{Fengnian_Xia_NC_2014, BAs_AM2018, Xiaolong_Xu_ACSAMI_2017, Yinan_Liu_Adv_Mater_2018, Liang_Li_ACSNano_2017, MiaoFeng_NC_2015, Liang_Li_AdvMater_2018, Shengxue_Yang_AFM_2018} For example, ultra-thin black phosphorous flake showed an in-plane anisotropic conductance reaching a ratio $\sigma_{a}/\sigma_{b}$ of about 1.5, which in principle can be a direction-sensitive sensor.\cite{Fengnian_Xia_NC_2014} ReS$_{2}$ was reported to be another candidate for anisotropic 2D field effect transistor, which exhibited a $\sigma_{a}$ of almost ten times the value of $\sigma_{b}$.\cite{MiaoFeng_NC_2015} Recent studies on GeP and GeAs$_{2}$ flakes also showed anisotropic behavior with anisotropic factors of $1.5 \sim 2$ for their conductance.\cite{Liang_Li_AdvMater_2018,Shengxue_Yang_AFM_2018}

   \begin{figure*}[ht!]
   \includegraphics[width=0.85\linewidth]{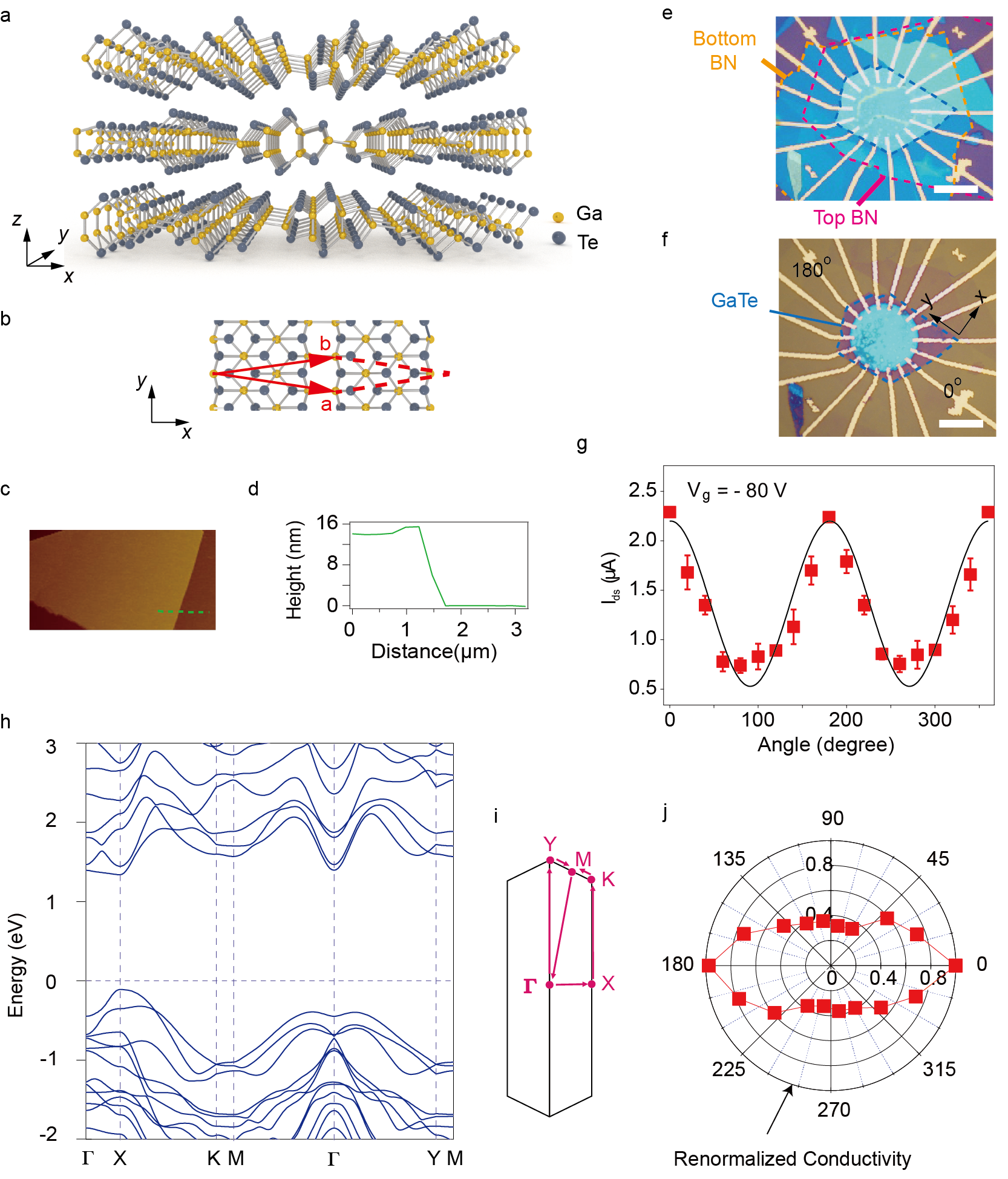}
   \caption{\textbf{Characterizations of ultra-thin GaTe.} (a) Schematics of GaTe crystal structure, with its in-plane unit cell illustrated in (b). (c) AFM image of a typical GaTe flake of about 14 nm in thickness, with its height profile along the green dashed line plotted in (d). (e) Optical image of a typical device (sample-S1) made of 14 nm GaTe flake encapsulated in h-BN. Electrodes are patterned with an angle interval of 20 degree. (f) Same device in (e) but patterned into a circular shape. Scale bars in (e) and (f) are 10 $\mu$m. and  (g) Source drain current at V$_{ds}$= 2V as a function of angle, with V$_{g}$=-80 V. The black solid line is an ellipsoidal fit. (h) Electronic band structure of monolayer GaTe, with the first Brillouin zone plotted in (i). (j) Same data in (g) plotted after re-normalization in a polar graph.
   }
   \end{figure*}

To date, in-plane anisotropic factors $\Gamma_{a}$ of electrical conductance in 2D materials are yet limited within one order of magnitude under ambient condition. It thus hinders future applications owing to a weak effective conductance difference between directions, and a larger anisotropic factor is highly desired. In this letter, we show an observation of giant anisotropic resistance (GAR) behaviour in few-layered $p$-type semiconducting GaTe. Among devices measured, electrical conductivity along $\textbf{\textit{x}}$ and $\textbf{\textit{y}}$ ($\textbf{\textit{x}}$ and $\textbf{\textit{y}}$ are defined in Fig.1) directions of ultra-thin GaTe reaches an order of 10$^{3}$ at gate voltages close to the valence band maximum (VBM), which can be further gate tuned down to less than 10, upon hole doping. We show detailed analysis and physical understandings on the GAR effect. Based on this, floating gate anisotropic memristors with directional multi-level outputs are demonstrated. Moreover, when measuring along fixed direction, the few-layered GaTe memristor exhibits On/Off ratio of 10$^{7}$ and retention time of 10$^{5}$ s, which is by far the best performance among memristors made of 2D materials. Our findings open new possibilities towards next generation nanoelectronics, such as artificial neuron network based on anisotropic memory arrays.

   \begin{figure*}[ht!]
   \includegraphics[width=0.85\linewidth]{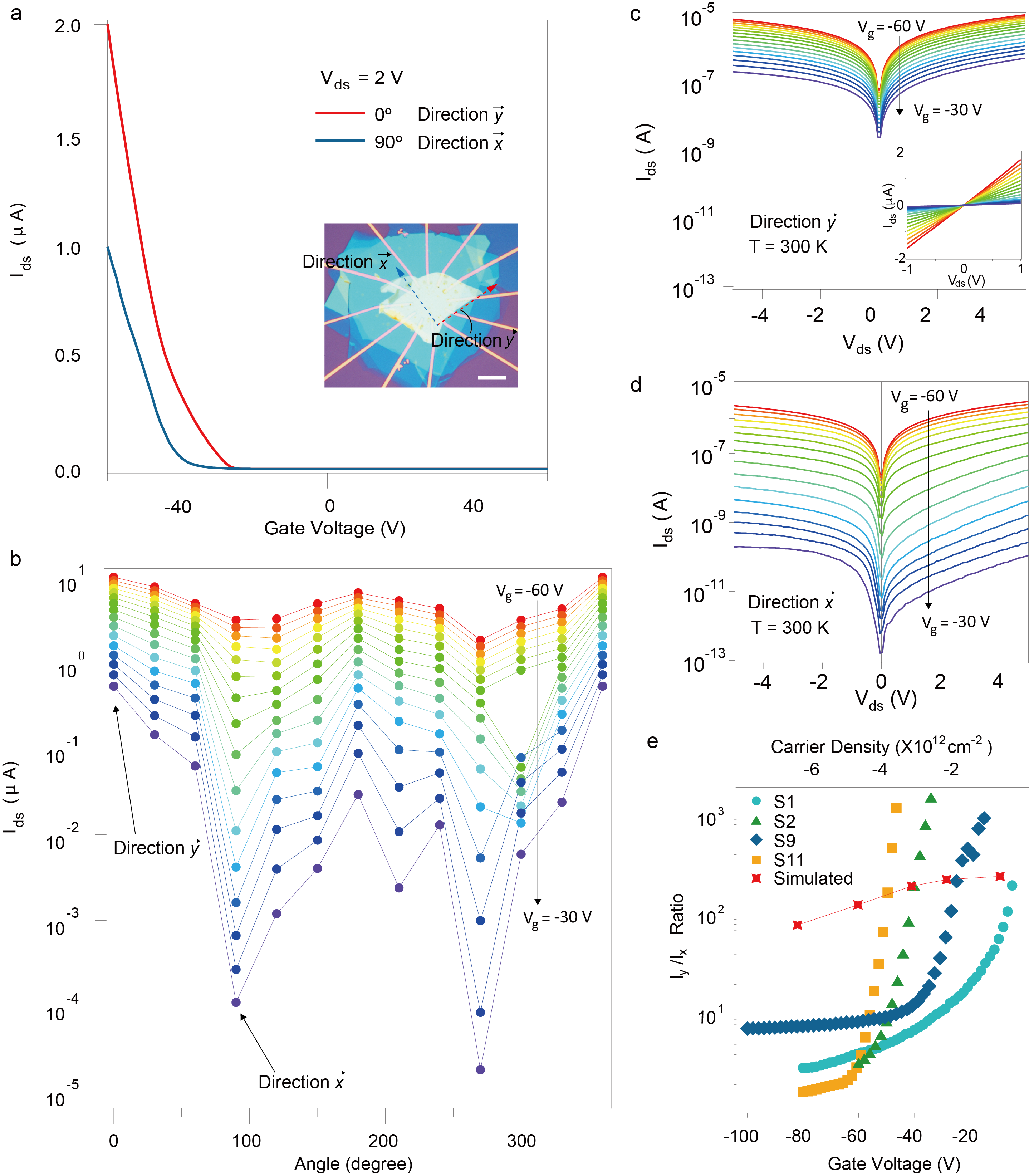}
   \caption{\textbf{Giant anisotropic resistance in ultra-thin GaTe.}  (a) Field effect curves of a typical GaTe device (Sample-S2) recorded along $\textbf{\textit{a}}$ and $\textbf{\textit{b}}$ directions, with the optical image of the device shown in the inset. Scale bar in the inset is 10 $\mu$m. (b) $I_{ds}$ of the same device measured along the 12 directions with a fixed $V_{ds}=5V$, and with the gate voltages swept from -60 V to -30 V. (c)-(d) $IV$ curves plotted in log scale at $V_{ds}= \pm 5V$ of the same device measured along $\textbf{\textit{y}}$ and $\textbf{\textit{x}}$ directions, respectively. The gate voltages were swept from -60 V to -30 V. Inset of (c) shows the corresponding data in linear scale. (e) The electrical maximum anisotropic ratio $I_{y}/I_{x}$ extracted from different samples (denoted as S1, S2, S9, and S11) as a function of gate voltage.}

   \label{fig:fig2}
   \end{figure*}

\section{Results}
\textbf{Characterizations of ultra-thin GaTe.} Bulk GaTe single crystals were prepared via flux method and were confirmed via x-ray diffraction (see Fig.S1). We mechanically exfoliated the bulk and deposited ultra-thin GaTe flakes onto 285 nm silicon oxide grown on heavily doped silicon wafers for  optical and electrical characterizations. It is known that GaTe has a layered structure with lattice symmetry of group C$_{2h}$ (Fig. 1a), similar to that of 1T$'$-MoTe$_{2}$. The unit cell for a single layer GaTe is indicated in Fig. 1b. Atomic Force Microscope (AFM) scan of a typical GaTe flake (14 nm in thickness) is shown in Fig. 1c-d. First, we characterized electrical transport of those flakes in air (Fig. S2-S6). However, without protection from air, source-drain current $I_{ds}$ of devices made by GaTe flakes are rather low, with maximum current of a few nano Amperes, in agreement with other reported results.\cite{Pingan_Hu_Nano_Research_2014} It is found that, without thermal annealing, few-layered GaTe devices are barely conducting, as shown in Fig. S2. In the following, we will focus on the devices made by ultra-thin GaTe flakes encapsulated by hexagonal boron nitride (h-BN) in glove box, in order to diminish damages from air (Fig. S7).

Polarized Raman can be used to determine crystal orientation in 2D flakes,\cite{Liang_Li_AdvMater_2018,Shengxi_Huang_ACS_Nano_2016} and also delivers information of phonon modes that are related to the lattice symmetry. Figure S8 shows Raman spectra of a typical GaTe flake, with the angle dependent intensity at Raman shift of 271.1 cm$^{-1}$ plotted in a polar figure overlaid on the optical image shown in the inset. A two-fold symmetry of the polar figure can be seen, with one of the polarized axis parallel to the exfoliated straight edge of the GaTe itself. Such a Raman-active mode is a A$_g$ mode in C$_{2h}$ point group. From Raman tensor analysis, the direction of maximal Raman intensity of such A$_g$ mode corresponds to the lattice direction with short lattice constant, defined as $\textbf{\textit{y}}$ axis in Fig. 1a. Raman mode at 164.6 cm$^{-1}$ is another good reference for determining the crystalline direction (see Fig. S9) ~\cite{Shengxi_Huang_ACS_Nano_2016}. Such a mode has a fourfold symmetry (B$_g$ mode) and therefore four-lope angular dependence, with each lope 45 degrees away from the polarization axis of the mode at 271.1 cm$^{-1}$. By comparing Raman anisotropy with the anisotropy of electrical conductivity (shown in the next section), it is known that $\textbf{\textit{y}}$ axis of GaTe layer corresponds to a maximum conductivity (Fig. S8-S9). By this means, one can pattern electrodes for electrical measurements just with the zero angle defined along with such straight edges ($\textbf{\textit{y}}$ direction).

As shown in Fig. 1e, nine pairs of electrodes (20$^{o}$ angle between each two electrodes) were deposited onto the h-BN/GaTe/h-BN stack sample S1 (see Methods), which is further patterned into a circular shape in Fig. 1f. As a result, the source drain current $I_{ds}$ along each pair of electrodes at gate voltage $V_{g} = -80$ V follows an ellipsoidal dependence of the testing angle, shown in Fig. 1g. It is rather straight forward that the maximum $I_{ds}$ flows in 0 $^{o}$, which is parallel to one of the straight edges, i.e., the $\textbf{\textit{y}}$ axis, as marked in Fig. 1f. Such two-fold ellipsoidal oscillation of in-plane anisotropic conductivity is also seen in a number of 2D materials.\cite{SnSe, Fengnian_Xia_NC_2014, BAs_AM2018,  MiaoFeng_NC_2015, Liang_Li_AdvMater_2018, Shengxue_Yang_AFM_2018}

To unveil the origin of the anisotropy of in-plane electrical conductivity found in ultra-thin GaTe, we performed first-principles electronic structure calculations and non-equilibrium Green's functional quantum transport with a simplified model on GaTe monolayer. Electronic band structure along high-symmetry line is shown in Fig. 1h, with the first Brillouin zone and high-symmetry points defined in Fig. 1i. A direct band gap of 1.5 eV locates at the X point, in agreement with photo luminescence measurement in Fig. S10. Thanks to the direct band gap, outstanding photo responses have been reported in GaTe devices.\cite{HeJun_ACSnano, Fucai_Liu_ACS_Nano_2014} What's very interesting is the strongly anisotropic band dispersion obviously seen along two perpendicular directions ($\Gamma$-X and X-K). At the VBM, band dispersion along $\Gamma$-X is much bigger than along X-K, which gives rise to strong anisotropic effective mass m$^{*}$ at VBM, as shown in Fig. S11 in the Suppl. Info. m$^{*}$ along X-K direction is about 10 times larger than that along $\Gamma$-X, which seems to lead to a better transport property along $\textbf{\textit{x}}$ axis than $\textbf{\textit{y}}$ axis. However, deformation potential E$_1$ (shown in Fig. S12) due to electron-phonon scattering shows an opposite trend to m$^*$ in both $\textbf{\textit{x}}$ and $\textbf{\textit{y}}$ axis, and dominates over the anisotropy of effective mass to give rise to the anisotropy of electrical conductivity $\sigma$ according to the deformation potential theory (Method part), $\sigma_y$ is about 5 times as big as $\sigma_x$ at low doping. When plotting experimental data in the renormalized conductivity polar figure in Fig. 1j, a two-fold symmetry of conductance can be clearly seen.

\begin{figure*}[ht!]
\includegraphics[width=0.85\linewidth]{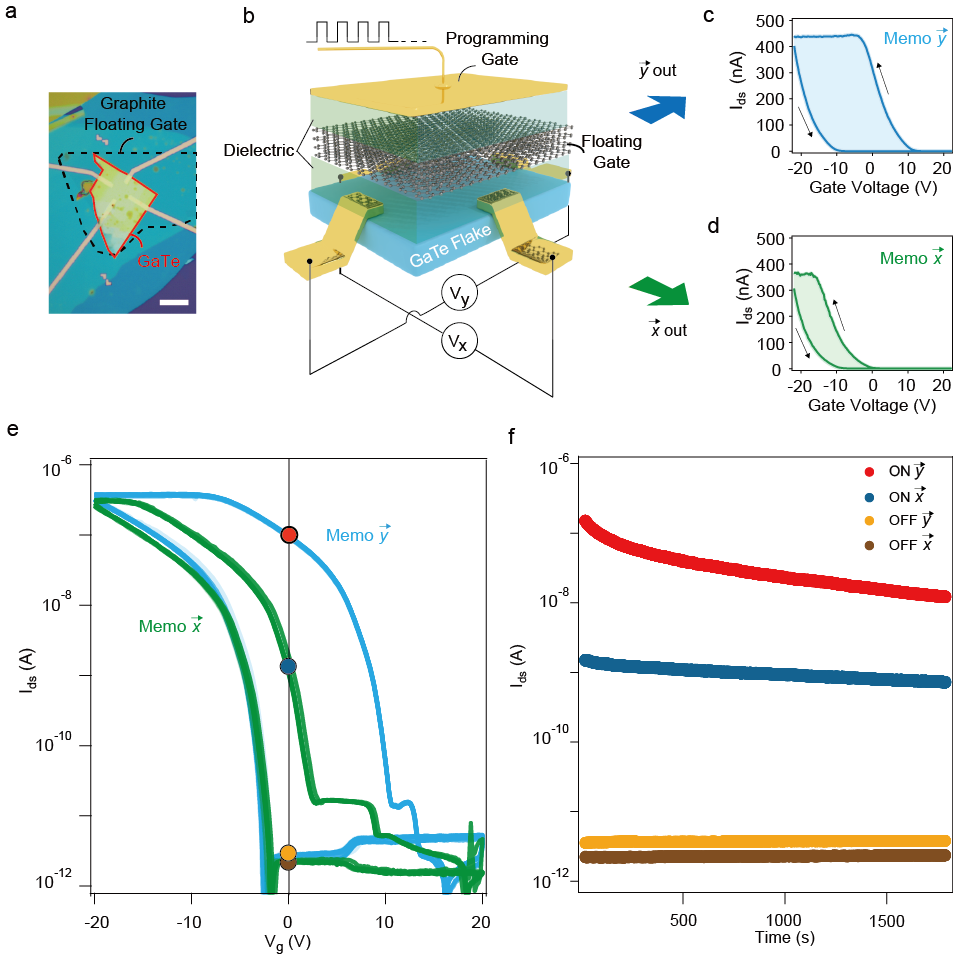}
\caption{\textbf{Anisotropic memristor based on GaTe floating gate memory.} (a)-(b) Optical image and art view of a typical h-BN/GaTe/h-BN device (Sample-S6) with a graphite floating gate. Scale bar in (a) is 10 $\mu$m. (c)-(d) Memory curves measured along $\textbf{\textit{x}}$ and $\textbf{\textit{y}}$ directions, respectively. (e) Same data in (c) and (d), plotted in log scales, with 10 repeated measurements (indicated by changing the gradient of the green and blue colours). (f) Retention time of memory at $V_{g}$=0 V, along $\textbf{\textit{y}}$ and $\textbf{\textit{x}}$ directions, with on and off positions indicated by the coloured circles in (e).}

\label{fig:fig4}
\end{figure*}

\begin{figure*}[ht!]
\includegraphics[width=0.85\linewidth]{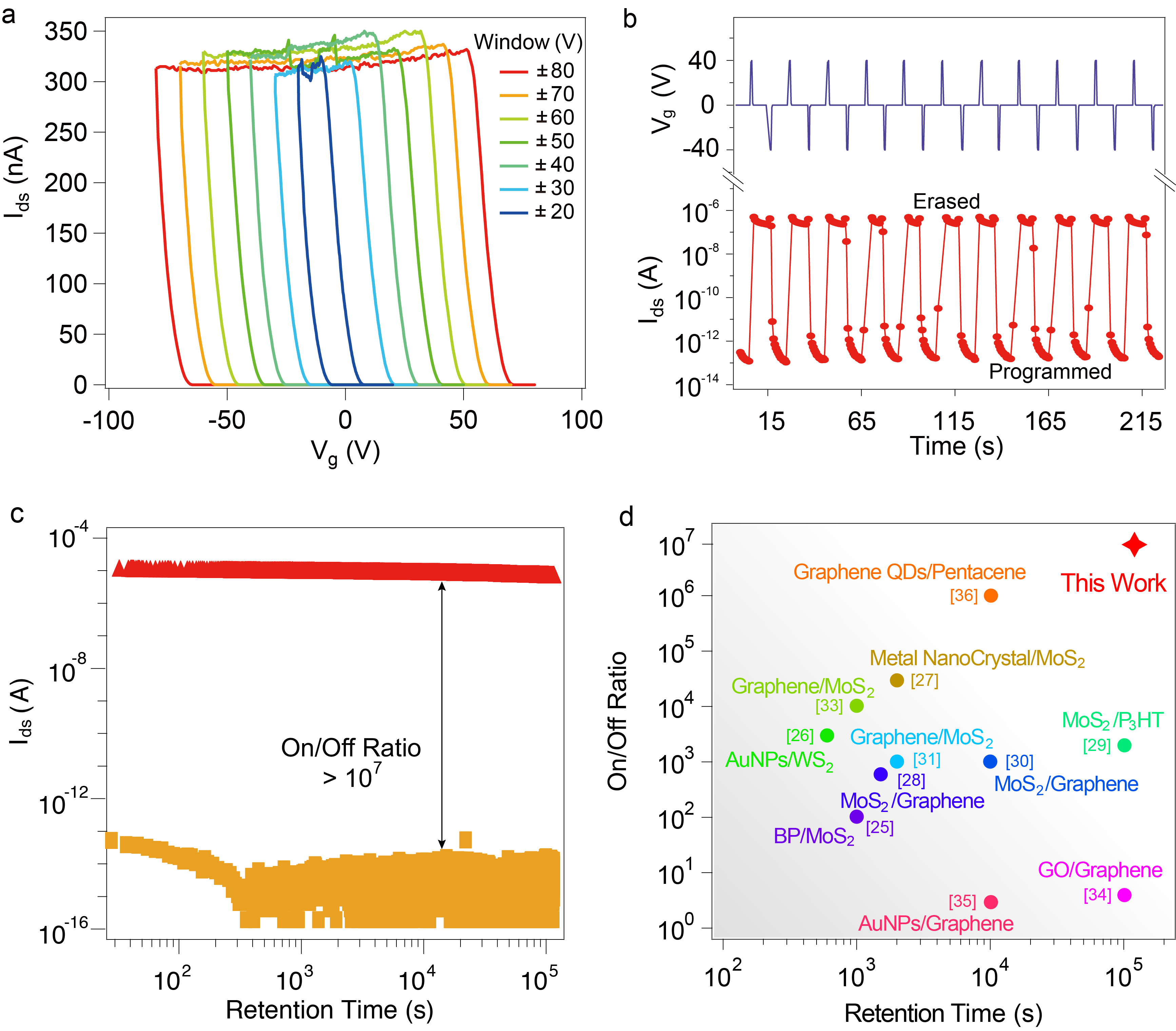}
\caption{\textbf{Comparison the floating gate memristor performance of GaTe with other 2D materials.} (a) Memory windows measured in $\textbf{\textit{y}}$ direction. (b) Demonstration of erasing and programming pulses measured in $\textbf{\textit{y}}$ direction. (c) Retention time test of the device at on and off states. (d) Summary of the performance of FGM made by 2D materials. Data in (a)-(b) are measured in sample S6, and data in (c)-(d) are measured in sample S4.}

\label{fig:fig5}
\end{figure*}

\textbf{Gate-tunable giant anisotropic resistance in ultra-thin GaTe.} The above mentioned two-fold in-plane symmetry of resistivity can be reproduced in multiple GaTe devices (Fig. S13-S14), and we performed a systematic study on the sample S2. As shown in Figure 2a, with the same GaTe flake (optical image of the device is shown in the inset of Fig. 2a), field effect curves, measured at source drain voltage $V_{ds}=2V$ along $\textbf{\textit{x}}$ and $\textbf{\textit{y}}$ directions, exhibit strong anisotropy, indicated by red and blue colours, respectively. $I_{ds}$ measured along the 12 directions with a fixed $V_{ds}=5$ V are plotted at $V_{g}=-60$ V, while an ellipsoidal oscillation at all angles with 2$\pi$ periodicity can be seen, shown in Fig. 2b. When the gate voltages were swept from -60 V to -30 V, the maximum-to-minimum $I_{ds}$ ratio of the ellipsoidal oscillation at each gate voltage varies largely. Note that similar gate tunable anisotropic resistance beahviour is also seen in bare GaTe device without BN encapsulation (Fig. S15-S16). 

Taking $\textbf{\textit{y}}$ (0 $^{o}$) and $\textbf{\textit{x}}$ (90 $^{o}$) directions for examples, closer look of $IV$ output curves in the range of $\pm 5$ V are shown in Fig. 2c-d. It is seen that at high bias, $I_{ds}$ for direction-$\textbf{\textit{y}}$ is varying within 10$^{-7}$ and 10$^{-5}$ A, while $I_{ds}$ for direction-$\textbf{\textit{x}}$ is varying within 10$^{-11}$ and 10$^{-6}$ A. We define the electrical maximum anisotropic ratio $\Gamma_{a}$ as $I_{y}/I_{x}$ for each fixed gate voltage. Strikingly, as extracted from Fig. 2b, $\Gamma_{a}$ as a function of gate voltage shows a gate-tunability from less than one order to as large as 10$^{3}$. This gate-tunable GAR effect was not found in previous studies. We fabricated various samples in similar configuration of h-BN encapsulated ultra-thin GaTe devices (Fig. S17-S19), with their $\Gamma_{a}$ plotted together in Fig. 2e. In general, anisotropic ratio in those devices can be gate tuned from a few times, to several hundreds or even  thousands folds, which is much higher than other 2D systems reported.\cite{BAs_AM2018,MiaoFeng_NC_2015,Fengnian_Xia_NC_2014,Yinan_Liu_Adv_Mater_2018,Gang_Qiu_Nano_Lett_2016}

Since $\Gamma_{a}$ in ultra-thin GaTe is rather weak at relatively higher hole doping (say, in the $V_{g}=-60$ V limit), and becomes significantly enhanced close to the VBM in the vicinity of $V_{g}=-30$ V, it more or less shows a better electron collimation at higher $\Gamma_{a}$ values, i.e., electrons tend to flow preferably along a certain direction at those conditions. Considering the deformation potential theory only applied to the band edge (VBM or CBM) and its inability to take care the gate-tunability on electrical transport,\cite{Shockley1950, Dong19} we calculate $IV$ curve and its gating dependence by combining density functional theory (DFT) calculation with the non-equilibrium Green's function (NEGF) method~\cite{Brandbyge02} (See method part), and show calculated field effect $IV$ data at $V_{ds}$=0.5 V along two perpendicular crystalline directions in Fig.S20. To avoid possible contact Schottky barrier, we use p-type heavily doped GaTe as source and drain. $IV$ curves along both $\textbf{\textit{x}}$ and $\textbf{\textit{y}}$ directions show a similar behavior to the experimental ones as given in Fig.2a. $I_{y,ds}$ at $V_{ds}$ = 0.5 V increases from 1.39 $\mu$A at $V_{g}$ = -9.1 V to 12.52 $\mu$A at $V_{g}$ = -82 V. More strikingly is the gate dependence of $I_{x,ds}$ which changes by two orders of magnitude, from 4.63 $\times$10$^{-3}$ $\mu$A at $V_{g}$=-9.1 V to 0.19 $\mu$A at $V_{g}$=-82 V. The calculated $I_y$/$I_x$ ratio, given in filled red squares in Fig. 2e, decreases with increasing gate voltage, from $\approx$300 at -9.1 V to $\approx$60 at -82 V, agreeing qualitatively with the experimental ratio but in a milder way than the experimental data. The discrepancy can be ascribed to that extrinsic effect in experiment such as nonlinear contact resistivity is not taken into account in the simulated device. Nevertheless, the gate-tunable GAR can still be intrinsically analysed from the calculated transmission coefficient, as shown in Fig.S21. At low gate voltage (-9.1 V), there is almost no $\textbf{\textit{x}}$-direction transmission channels in the scattering region between source energy level E$_{L}$ and drain energy E$_{R}$, while a sizable $\textbf{\textit{y}}$-direction transmission is observed, giving rise to a large anisotropic ratio $\Gamma_{a}$ at low gate voltages. In contrast, at high gate voltage (-82 V), non-zero transmission coefficient in channel material appears with comparable total transmission in both $\textbf{\textit{x}}$ and $\textbf{\textit{y}}$ directions, greatly suppressing GAR in GaTe.

\bigskip

\textbf{Anisotropic memristor based on GaTe floating gate memory.} It is of great importance to demonstrate conceptual devices taking advantage of the GAR effect in ultra-thin GaTe. In the following, we will discuss a prototype anisotropic memory based on ultra-thin GaTe. Using the vdW assembly method (see Methods), a few-layered graphene was equipped in addition to the h-BN/GaTe/h-BN sandwich, forming an architecture of floating gate memory (FGM). Optical image and art view of such typical devices are shown in Fig. 3a-b.

Interestingly, when sweeping within the same gate window, the hysteresis memory curves along $\textbf{\textit{y}}$ and $\textbf{\textit{x}}$ directions vary largely, with a clear discrepancy of memory window size, shown in Fig. 3c-d. This result is more pronounced in a log-scale when plotted together in Fig. 3e. One can see that after an operation of programming (sweep the $V_{g}$ from 0 V to -20 V, and back to 0 V), the device stays in an `Off' state in both directions. While an erase operation (sweep the $V_{g}$ from 0 V to 20 V, and back to 0V), the device stays in two different `On' states in $\textbf{\textit{y}}$ and $\textbf{\textit{x}}$ directions, as indicated by coloured circles in Fig. 3e, and their corresponding retention in Fig. 3f. It thus makes the device an anisotropic memristor, i.e., one single operation of programming-erasing will generate two sets of data in the two directions, making the device conceptually compatible with direction-sensitive data storage.

\section{Discussion}

To verify the performance of ultra-thin GaTe with graphite floating gate memristor along fixed direction, we recorded polarization gate voltages for different ranges from $\pm$ 20 V to $\pm$ 80 V in direction-$\textbf{\textit{y}}$, shown in Fig. 4a (see also Fig. S22-S23). It can be seen that the GaTe memristor shows rather stable erased state in all gate ranges. Once fed with the stimulation of pulsed $V_{g}$ of $\pm$ 40 V, the device can be readily erased and programmed as illustrated in Fig. 4b.

In the following, we compare the GaTe FGM with the state-of-the-art memristors fabricated from other 2D materials.\cite{Dong_Li_AFM_2015,Fan_Gong_AFM_2016,Jingli_Wang_Samll_2014,Min_Sup_Choi_NC_2013,Minji_Kang_Nanoscale_2014,Quoc_An_Vu_NC_2016,Quoc_A_V_Adv_Mater_2018,Simone_Bertolazzi_ACS_Nano_2013,Sukjae_Jang_Small_2014,Sukjae_Jang_Nanoscale_2014,Yongsung_Ji_Nanotech_2016} Given that retention of the erased and/or programmed states is one of the most important parameter of such FGM, we recorded along direction-$\textbf{\textit{y}}$ the on state (by a positive gate pulse), and the off state (by a negative gate pulse), respectively. Among GaTe FGM devices tested, the best performance (measured from sample-S4 as shown in Fig. 4c) exhibits an on/off ratio exceeding 10$^{7}$ and the attenuation of the on/off ratio is less than 5 $\%$ in a retention time of 10$^{5}$ s (detailed memory characterizations of sample S4 are shown in Fig. S24). By summarizing the state-of-the-art performance of FGM based on 2D materials, it is found that, our h-BN/GaTe/h-BN with graphite floating gate memory shows up to now the record on/off ratio and retention time, as shown in Fig. 4d.


In conclusion, we have discovered that ultra-thin GaTe encapsulated by h-BN devices show a two-fold symmetry electrical conductance oscillations along different in-plane directions, with their maximum anisotropic ratio $I_{y}/I_{x}$ gate tunable in the range from less than 10 to a few thousands. This giant anisotropic resistance (GAR) arises from the in-plane anisotropic transmission coefficients, and the anisotropic ratio can be changed by several orders of magnitude mainly due to the sensitivity of $\textbf{\textit{x}}$-direction conduction channel to gating. Based on this GAR effect, we devised an anisotropic GaTe memory with a vdW assembled graphite floating gate. The prototype memory devices show unprecedented direction-sensitive multiple-level output memristor behaviour when measured from different directions. Moreover, fixed direction measurements suggest that the GaTe FGMs show record performances in terms of both on/off ratio and retention time among the state-of-the-art FGM based on 2D materials. Using this unit cell of GaTe anisotropic field effect transistor as a building block, it can be further expanded into many possible applications such as neuronal computation with GaTe FGM neural network arrays.

\section{Methods}

Single crystal GaTe was prepared via the self-flux method. Raw bulk material with stoichiometric ratio of Ga (purity 99.999$\%$): Te (purity 99.9$\%$)=48.67:51.33 were mixed and kept at 880 $^{o}$C for 5 h. The mixture was then cooled at the rate of 1.5 $^{o}$C h$^{-1}$, followed by a natural cooling process. The h-BN (crystals from HQ Graphene) encapsulated GaTe devices are fabricated using the dry-transfer methods (Fig. S25).\cite{Wang_Zhi_Nat_Nano_2018_CGT}

The electronic band structure in this work is calculated by using the first-principles density functional theory as implemented in the \textsc{VASP} code\cite{VASP}. Projector augmented wave (PAW) pseudopotentials~\cite{PAWPseudo} and the Perdew-Burke-Ernzerhof (PBE)~\cite{PBE} functional are respectively used to describe electron-ion interaction and electronic exchange-correlation interaction. We adopt $500$~eV as the electronic kinetic energy cutoff for the plane-wave basis and $10^{-6}$~eV as the criterion for reaching self-consistency. The Brillouin zone (BZ) of the primitive unit cell (12 atoms per cell) is sampled by $20{\times}20{\times}1$ $k$-points~\cite{Monkhorst-Pack76}. Rectangle supercell (24 atoms per cell) is used to calculate anisotropic electrical conductivity and the BZ is sampled by $2{\times}6{\times}1$ $k$-points.

The electrical conductivity $\sigma$ is evaluated based on $\sigma$ = ne$\mu$, in which carrier mobility $\mu$ is calculated based upon the deformation potential theory with formula~\cite{Shockley1950,Dong19}
\begin{equation}
    \mu = \frac{e\hbar^{3}C_{2D}}{k_{B}Tm^{*}m_{d}E_{1}^{2}},
\label{eq3}
\end{equation}
where $\hbar$ is the reduced Planck constant, $C_{\textrm{2D}}$ is the elastic modulus derived from $(E-E_{0})/A_{0} = C\varepsilon^{2}$/2 ($E$, $E_0$, $A_0$, $C$, and $\varepsilon$ denote, respectively, the total energy, the total energy at equilibrium, the area of the 2D unit cell at equilibrium, elastic constant, and the lattice strain), $k_{\textrm{B}}$ is the Boltzmann constant, $T$ is temperature, $m^*$ is the effective mass in the transport direction, $m_{d}$ is the averaged effective mass defined by $m_{d} = \sqrt{m_{x}^{*}m_{y}^{*}}$ and $E_{1}$ is the deformation potential constant of the valence-band maximum (VBM) along the transport direction. $E_{1}$ defined by $E_{1} = \Delta V/\varepsilon$ with $\Delta V$ as the energy change of the VBM under small strain $\varepsilon$.

$I-V$ curve of the GaTe transistor is simulated using the DFT coupled with the non-equilibrium Green's function (NEGF) method~\cite{Brandbyge02}, as implemented in the ATK package~\cite{atk18}. The structure of GaTe transistor is shown in Fig. S20 (a). We employ the Dirichlet boundary condition to ensure the charge neutrality in the source and the drain region. The temperature is set to 300 K. The density mesh cutoff is set to 80 Hartrees. The Monkhorst-Pack k-point meshes for device along the x and y direction are sampled with 40 $\times$ 6 $\times$ 1 and 114 $\times$ 6 $\times$ 1, respectively. Electron-ion interaction is treated by SG15 Optimized Norm-Conserving Vanderbilt (ONCV) pseudopotentials~\cite{PhysRevB.88.085117,Schlipf15}. The transmission coefficient at energy E averaged over 31 k-points perpendicular to the transport direction. The generalized gradient approximation (GGA) is adopted for the electronic exchange-correction functional. The length of scattering region is 9.530 nm and 9.676 nm for transistor along x and y direction, respectively, both of which are sufficiently large to avoid interaction between source and drain.

\section{\label{sec:level1}ACKNOWLEDGEMENT}
This work is supported by the National Key R$\&$D Program of China (2017YFA0206302), and is supported by the National Natural Science Foundation of China (NSFC) with Grant 11504385 and 51627801. T.Yang acknowledges supports from the Major Program of Aerospace Advanced Manufacturing Technology Research Foundation NSFC and CASC, China (No. U1537204). L.T. Xiao acknowledges support from the National Key R$\&$D Program of China (2017YFA0304203). H.G. acknowledges NSFC Grant No. 51702146. Z. Han acknowledges the support from the Program of State Key Laboratory of Quantum Optics and Quantum Optics Devices (No. KF201816).

\section{Author contributions}
Z.H., Z.Z. and T.Y. conceived the experiment and supervised the overall project. H.W., X.S., J.C. fabricated the devices. H.W., M.C., X.S., X.L., Y.Y. and Z.H. carried out electrical transport measurements; M.Z., C.Q., X.Z., J.Z., S.C. and J.C. carried out optical measurements; Y.W., B.D. and T.Y. conducted the theoretical simulations. J.H. and H.G. did the simulations on polarized Raman. D.S. contributed in sample fabrications. L.Z. and W.L. grew the GaTe bulk crystals. The manuscript was written by Z.H., T.Y., and H.W. with discussion and inputs from all authors.


\begin{thebibliography}{10}
\expandafter\ifx\csname url\endcsname\relax
  \def\url#1{\texttt{#1}}\fi
\expandafter\ifx\csname urlprefix\endcsname\relax\def\urlprefix{URL }\fi
\providecommand{\bibinfo}[2]{#2}
\providecommand{\eprint}[2][]{\url{#2}}

\bibitem{Fengnian_Xia_NC_2014}
\bibinfo{author}{Xia, F.}, \bibinfo{author}{Wang, H.} \& \bibinfo{author}{Jia,
  Y.}
\newblock \bibinfo{title}{Rediscovering black phosphorus as an anisotropic
  layered material for optoelectronics and electronics}.
\newblock \emph{\bibinfo{journal}{Nat. Commun.}} \textbf{\bibinfo{volume}{5}},
  \bibinfo{pages}{4458} (\bibinfo{year}{2014}).

\bibitem{BAs_AM2018}
\bibinfo{author}{Chen, Y.} \emph{et~al.}
\newblock \bibinfo{title}{Black arsenic: A layered semiconductor with extreme
  in-plane anisotropy}.
\newblock \emph{\bibinfo{journal}{Adv. Mater.}} \textbf{\bibinfo{volume}{30}},
  \bibinfo{pages}{1800754} (\bibinfo{year}{2018}).

\bibitem{Xiaolong_Xu_ACSAMI_2017}
\bibinfo{author}{Xu, X.} \emph{et~al.}
\newblock \bibinfo{title}{In-plane anisotropies of polarized raman response and
  electrical conductivity in layered tin selenide}.
\newblock \emph{\bibinfo{journal}{ACS Appl. Mater. Interfaces}}
  \textbf{\bibinfo{volume}{9}}, \bibinfo{pages}{12601--12607}
  (\bibinfo{year}{2017}).

\bibitem{Yinan_Liu_Adv_Mater_2018}
\bibinfo{author}{Liu, Y.} \emph{et~al.}
\newblock \bibinfo{title}{Raman signatures of broken inversion symmetry and
  in-plane anisotropy in type-{II} {W}eyl semmimetal candidate
  {T}a{I}r{T}e$_{4}$}.
\newblock \emph{\bibinfo{journal}{Adv. Mater.}} \textbf{\bibinfo{volume}{30}},
  \bibinfo{pages}{1706402} (\bibinfo{year}{2018}).

\bibitem{Liang_Li_ACSNano_2017}
\bibinfo{author}{Li, L.} \emph{et~al.}
\newblock \bibinfo{title}{Strong in-plane anisotropies of optical and
  electrical response in layered dimetal chalcogenide}.
\newblock \emph{\bibinfo{journal}{ACS Nano}} \textbf{\bibinfo{volume}{11}},
  \bibinfo{pages}{10264--10272} (\bibinfo{year}{2017}).

\bibitem{MiaoFeng_NC_2015}
\bibinfo{author}{Liu, E.} \emph{et~al.}
\newblock \bibinfo{title}{Integrated digital inverters based on two-dimensional
  anisotropic {R}e{S}$_{2}$ field-effect transistors}.
\newblock \emph{\bibinfo{journal}{Nat. Commun.}} \textbf{\bibinfo{volume}{6}},
  \bibinfo{pages}{6991} (\bibinfo{year}{2015}).

\bibitem{Liang_Li_AdvMater_2018}
\bibinfo{author}{Li, L.} \emph{et~al.}
\newblock \bibinfo{title}{2{D} {G}e{P}: An unexploited low-symmetry
  semiconductor with strong in-plane anisotropy}.
\newblock \emph{\bibinfo{journal}{Adv. Mater.}} \textbf{\bibinfo{volume}{30}},
  \bibinfo{pages}{1706771} (\bibinfo{year}{2018}).

\bibitem{Shengxue_Yang_AFM_2018}
\bibinfo{author}{Yang, S.} \emph{et~al.}
\newblock \bibinfo{title}{Highly in-plane optical and electrical anisotropy of
  2{D} germanium arsenide}.
\newblock \emph{\bibinfo{journal}{Adv. Funct. Mater.}}
  \textbf{\bibinfo{volume}{28}}, \bibinfo{pages}{1707379}
  (\bibinfo{year}{2018}).

\bibitem{Nature1939}
\bibinfo{author}{Krishnan, K.~S.} \& \bibinfo{author}{Ganguli, N.}
\newblock \bibinfo{title}{Large anisotropy of the electrical conductivity of
  graphite}.
\newblock \emph{\bibinfo{journal}{Nature}} \textbf{\bibinfo{volume}{144}},
  \bibinfo{pages}{667} (\bibinfo{year}{1939}).

\bibitem{MgB2}
\bibinfo{author}{Eltsev, Y.} \emph{et~al.}
\newblock \bibinfo{title}{Anisotropic resistivity and hall effect in
  {M}g{B}$_{2}$ single crystals}.
\newblock \emph{\bibinfo{journal}{Phys. Rev. B}} \textbf{\bibinfo{volume}{66}},
  \bibinfo{pages}{180504(R)} (\bibinfo{year}{2002}).

\bibitem{WTe2_JWang}
\bibinfo{author}{Zhao, Y.}
\newblock \bibinfo{title}{Anisotropic magnetictransport and exotic longitudinal
  linear magnetoresistance in {WT}e$_{2}$ crystals}.
\newblock \emph{\bibinfo{journal}{Phys. Rev. B}} \textbf{\bibinfo{volume}{92}},
  \bibinfo{pages}{041104(R)} (\bibinfo{year}{2015}).

\bibitem{Pingan_Hu_Nano_Research_2014}
\bibinfo{author}{Hu, P.} \emph{et~al.}
\newblock \bibinfo{title}{Highly sensitive phototransistors based on
  two-dimensional {G}a{T}e nanosheets with direct bandgap}.
\newblock \emph{\bibinfo{journal}{Nano Research}} \textbf{\bibinfo{volume}{7}},
  \bibinfo{pages}{694--703} (\bibinfo{year}{2014}).

\bibitem{Shengxi_Huang_ACS_Nano_2016}
\bibinfo{author}{Huang, S.} \emph{et~al.}
\newblock \bibinfo{title}{In-plane optical anisotropy of layered gallium
  telluride}.
\newblock \emph{\bibinfo{journal}{ACS Nano}} \textbf{\bibinfo{volume}{10}},
  \bibinfo{pages}{8964--8972} (\bibinfo{year}{2016}).

\bibitem{SnSe}
\bibinfo{author}{Yang, S.~X.} \emph{et~al.}
\newblock \bibinfo{title}{Highly-anisotropic optical and electrical properties
  in layered {S}n{S}e}.
\newblock \emph{\bibinfo{journal}{Nano Research}}
  \textbf{\bibinfo{volume}{11}}, \bibinfo{pages}{554--564}
  (\bibinfo{year}{2018}).

\bibitem{HeJun_ACSnano}
\bibinfo{author}{Wang, Z.} \emph{et~al.}
\newblock \bibinfo{title}{Role of {G}a vancancy on a multilayer {G}a{T}e
  phototransistor}.
\newblock \emph{\bibinfo{journal}{ACS Nano}} \textbf{\bibinfo{volume}{8}},
  \bibinfo{pages}{4859--4865} (\bibinfo{year}{2014}).

\bibitem{Fucai_Liu_ACS_Nano_2014}
\bibinfo{author}{Liu, F.}, \bibinfo{author}{Shimotani, H.},
  \bibinfo{author}{Shang, H.} \& \bibinfo{author}{Kanagasekaran, T. e.~a.}
\newblock \bibinfo{title}{High-sensitivity photodetectors based on multilayer
  {G}a{T}e flakes}.
\newblock \emph{\bibinfo{journal}{ACS Nano}} \textbf{\bibinfo{volume}{8}},
  \bibinfo{pages}{752--760} (\bibinfo{year}{2014}).

\bibitem{Gang_Qiu_Nano_Lett_2016}
\bibinfo{author}{Qiu, G.} \emph{et~al.}
\newblock \bibinfo{title}{Observation of optical and electrical in-plane
  anisotropy in highmobility few-layer {Z}r{T}e$_{5}$}.
\newblock \emph{\bibinfo{journal}{Nano Letters}} \textbf{\bibinfo{volume}{16}},
  \bibinfo{pages}{7364--7369} (\bibinfo{year}{2016}).

\bibitem{Shockley1950}
\bibinfo{author}{Bardeen, J.} \& \bibinfo{author}{Shockley, W.}
\newblock \bibinfo{title}{Deformation potentials and mobilities in non-polar
  crystals}.
\newblock \emph{\bibinfo{journal}{Phys. Rev.}} \textbf{\bibinfo{volume}{80}},
  \bibinfo{pages}{72--80} (\bibinfo{year}{1950}).
\newblock \urlprefix\url{https://link.aps.org/doi/10.1103/PhysRev.80.72}.

\bibitem{Dong19}
\bibinfo{author}{Dong, B.} \emph{et~al.}
\newblock \bibinfo{title}{New two-dimensional phase of tin chalcogenides:
  candidates for high-performance thermoelectric materials}.
\newblock \emph{\bibinfo{journal}{Phys. Rev. Materials}}
  (\bibinfo{year}{2019}).

\bibitem{Brandbyge02}
\bibinfo{author}{Brandbyge, M.}, \bibinfo{author}{Mozos, J.-L.},
  \bibinfo{author}{Ordej\'on, P.}, \bibinfo{author}{Taylor, J.} \&
  \bibinfo{author}{Stokbro, K.}
\newblock \bibinfo{title}{Density-functional method for nonequilibrium electron
  transport}.
\newblock \emph{\bibinfo{journal}{Phys. Rev. B}} \textbf{\bibinfo{volume}{65}},
  \bibinfo{pages}{165401} (\bibinfo{year}{2002}).
\newblock \urlprefix\url{https://link.aps.org/doi/10.1103/PhysRevB.65.165401}.

\bibitem{Dong_Li_AFM_2015}
\bibinfo{author}{Li, D.} \emph{et~al.}
\newblock \bibinfo{title}{Nonvolatile floating-gate memories based on stacked
  black phosphorus-boron nitride-{M}o{S}$_{2}$ heterostructures}.
\newblock \emph{\bibinfo{journal}{Adv. Funct. Mater.}}
  \textbf{\bibinfo{volume}{25}}, \bibinfo{pages}{7360--7365}
  (\bibinfo{year}{2015}).

\bibitem{Fan_Gong_AFM_2016}
\bibinfo{author}{Gong, F.} \emph{et~al.}
\newblock \bibinfo{title}{High-sensitivity floating-gate phototransistors based
  on {WS}$_{2}$ and {M}o{S}$_{2}$}.
\newblock \emph{\bibinfo{journal}{Adv. Funct. Mater.}}
  \textbf{\bibinfo{volume}{26}}, \bibinfo{pages}{6084--6090}
  (\bibinfo{year}{2016}).

\bibitem{Jingli_Wang_Samll_2014}
\bibinfo{author}{Wang, J.} \emph{et~al.}
\newblock \bibinfo{title}{Floating gate memory-based monolayer {M}o{S}$_{2}$
  transistor with metal nanocrystals embedded in the gate dielectrics}.
\newblock \emph{\bibinfo{journal}{Samll}} \textbf{\bibinfo{volume}{11}},
  \bibinfo{pages}{208--213} (\bibinfo{year}{2015}).

\bibitem{Min_Sup_Choi_NC_2013}
\bibinfo{author}{Choi, M.~S.} \emph{et~al.}
\newblock \bibinfo{title}{Controlled charge trapping by molybdenum disulphide
  and graphene in ultrathin heterostructured memory devices}.
\newblock \emph{\bibinfo{journal}{Nat. Commun.}} \textbf{\bibinfo{volume}{4}},
  \bibinfo{pages}{1624} (\bibinfo{year}{2013}).

\bibitem{Minji_Kang_Nanoscale_2014}
\bibinfo{author}{Kang, M.} \emph{et~al.}
\newblock \bibinfo{title}{Stable charge storing in two-dimensional
  {M}o{S}$_{2}$ nanoflake floating gates for multilevel organic flash memory}.
\newblock \emph{\bibinfo{journal}{Nanoscale}} \textbf{\bibinfo{volume}{6}},
  \bibinfo{pages}{12315} (\bibinfo{year}{2014}).

\bibitem{Quoc_An_Vu_NC_2016}
\bibinfo{author}{Vu, Q.~A.} \emph{et~al.}
\newblock \bibinfo{title}{Two-terminal floating-gate memory with van der waals
  heterostructures for ultrahigh on/off ratio}.
\newblock \emph{\bibinfo{journal}{Nat. Commun.}} \textbf{\bibinfo{volume}{7}},
  \bibinfo{pages}{12725} (\bibinfo{year}{2016}).

\bibitem{Quoc_A_V_Adv_Mater_2018}
\bibinfo{author}{Vu, Q.~A.} \emph{et~al.}
\newblock \bibinfo{title}{A high-on/off-ratio floating-gate memristor array on
  a flexible substrate via {CVD}-grown large-area 2{D} layer stacking}.
\newblock \emph{\bibinfo{journal}{Adv. Mater.}} \textbf{\bibinfo{volume}{29}},
  \bibinfo{pages}{1703363} (\bibinfo{year}{2017}).

\bibitem{Simone_Bertolazzi_ACS_Nano_2013}
\bibinfo{author}{Bertolazzi, S.}, \bibinfo{author}{Krasnozhon, D.} \&
  \bibinfo{author}{Kis, A.}
\newblock \bibinfo{title}{Nonvolatile memory cells based on
  {M}o{S}$_{2}$/graphene heterostructures}.
\newblock \emph{\bibinfo{journal}{ACS Nano}} \textbf{\bibinfo{volume}{7}},
  \bibinfo{pages}{3246--3252} (\bibinfo{year}{2013}).

\bibitem{Sukjae_Jang_Small_2014}
\bibinfo{author}{Jang, S.}, \bibinfo{author}{Hwang, E.}, \bibinfo{author}{Lee,
  J.~H.}, \bibinfo{author}{Park, H.~S.} \& \bibinfo{author}{Cho, J.~H.}
\newblock \bibinfo{title}{Graphene-graphene oxide floating gate transistor
  memory}.
\newblock \emph{\bibinfo{journal}{Small}} \textbf{\bibinfo{volume}{DOI:
  10.1002/smll.201401017}} (\bibinfo{year}{2014}).

\bibitem{Sukjae_Jang_Nanoscale_2014}
\bibinfo{author}{Jang, S.}, \bibinfo{author}{Hwang, E.} \&
  \bibinfo{author}{Cho, J.~H.}
\newblock \bibinfo{title}{Graphene nano-floating gate transistor memory on
  plastic}.
\newblock \emph{\bibinfo{journal}{Nanoscale}} \textbf{\bibinfo{volume}{6}},
  \bibinfo{pages}{15286} (\bibinfo{year}{2014}).

\bibitem{Yongsung_Ji_Nanotech_2016}
\bibinfo{author}{Ji, Y.} \emph{et~al.}
\newblock \bibinfo{title}{Graphene quantum dots as a highly efficient
  solution-processed charge trapping medium for organic nano-floating gate
  memory}.
\newblock \emph{\bibinfo{journal}{Nanotechnology}}
  \textbf{\bibinfo{volume}{27}}, \bibinfo{pages}{145204}
  (\bibinfo{year}{2016}).

\bibitem{Wang_Zhi_Nat_Nano_2018_CGT}
\bibinfo{author}{Wang, Z.} \emph{et~al.}
\newblock \bibinfo{title}{Electric-field control of magnetism in a few-layered
  van der waals ferromagnetic semiconductor}.
\newblock \emph{\bibinfo{journal}{Nature Nanotechnology}}
  \textbf{\bibinfo{volume}{13}}, \bibinfo{pages}{554--559}
  (\bibinfo{year}{2018}).

\bibitem{VASP}
\bibinfo{author}{Kresse, G.} \& \bibinfo{author}{Furthm\"uller, J.}
\newblock \bibinfo{title}{Efficient iterative schemes for \textit{ab initio}
  total-energy calculations using a plane-wave basis set}.
\newblock \emph{\bibinfo{journal}{Phys. Rev. B}} \textbf{\bibinfo{volume}{54}},
  \bibinfo{pages}{11169--11186} (\bibinfo{year}{1996}).

\bibitem{PAWPseudo}
\bibinfo{author}{Kresse, G.} \& \bibinfo{author}{Joubert, D.}
\newblock \bibinfo{title}{From ultrasoft pseudopotentials to the projector
  augmented-wave method}.
\newblock \emph{\bibinfo{journal}{Phys. Rev. B}} \textbf{\bibinfo{volume}{59}},
  \bibinfo{pages}{1758--1775} (\bibinfo{year}{1999}).
\newblock \urlprefix\url{http://link.aps.org/doi/10.1103/PhysRevB.59.1758}.

\bibitem{PBE}
\bibinfo{author}{Perdew, J.~P.}, \bibinfo{author}{Burke, K.} \&
  \bibinfo{author}{Ernzerhof, M.}
\newblock \bibinfo{title}{Generalized gradient approximation made simple}.
\newblock \emph{\bibinfo{journal}{Phys. Rev. Lett.}}
  \textbf{\bibinfo{volume}{77}}, \bibinfo{pages}{3865--3868}
  (\bibinfo{year}{1996}).
\newblock \urlprefix\url{http://link.aps.org/doi/10.1103/PhysRevLett.77.3865}.

\bibitem{Monkhorst-Pack76}
\bibinfo{author}{Monkhorst, H.~J.} \& \bibinfo{author}{Pack, J.~D.}
\newblock \bibinfo{title}{Special points for brillouin-zone integrations}.
\newblock \emph{\bibinfo{journal}{Phys. Rev. B}} \textbf{\bibinfo{volume}{13}},
  \bibinfo{pages}{5188--5192} (\bibinfo{year}{1976}).

\bibitem{atk18}
\bibinfo{note}{QuantumFATK version 2018.06, Synopsys QuantumWise A/S
  (www.quantumwise.com).}

\bibitem{PhysRevB.88.085117}
\bibinfo{author}{Hamann, D.~R.}
\newblock \bibinfo{title}{Optimized norm-conserving vanderbilt
  pseudopotentials}.
\newblock \emph{\bibinfo{journal}{Phys. Rev. B}} \textbf{\bibinfo{volume}{88}},
  \bibinfo{pages}{085117} (\bibinfo{year}{2013}).
\newblock \urlprefix\url{https://link.aps.org/doi/10.1103/PhysRevB.88.085117}.

\bibitem{Schlipf15}
\bibinfo{author}{Schlipf, M.} \& \bibinfo{author}{Gygi, F.}
\newblock \bibinfo{title}{Optimization algorithm for the generation of oncv
  pseudopotentials}.
\newblock \emph{\bibinfo{journal}{Computer Physics Communications}}
  \textbf{\bibinfo{volume}{196}}, \bibinfo{pages}{36--44}
  (\bibinfo{year}{2015}).
\newblock
  \urlprefix\url{http://www.sciencedirect.com/science/article/pii/S0010465515001897}.

\end{thebibliography}
\end{document}